\documentstyle[11pt,paspconf,epsf]{article}

\begin{document}

\title{The ESO Nearby Abell Cluster Survey: Properties of the Galaxies}

\author{P. Katgert}
\affil{Leiden Observatory}
\author{C.Adami}
\affil{Lab. d'Astronomie Spatiale, Marseille}
\author{P. de Theije}
\affil{Leiden Observatory}
\author{A.Mazure}
\affil{Lab. d'Astronomie Spatiale, Marseille}
\author{T. Thomas}
\affil{Leiden Observatory}
\author{A. Biviano}
\affil{Osservatorio Astronomico di Trieste}





\begin{abstract}

In this paper we summarize the main properties of the ENACS (the ESO
Nearby Abell Cluster Survey), which was one of the ESO Keyprogrammes
carried out in the first half of the 1990's, and we discuss some
recent work on the properties of the galaxies in the rich clusters in
the ENACS sample. We stress the importance of the ENACS sample as a
local, volume-limited sample of rich ACO clusters, which in many
respects can serve as a local calibration of the properties of
optically selected rich clusters. However, by itself the ENACS has
also provided several useful insights into the properties of a sample
of rich clusters, and of the galaxies in them.

Here, we describe recent results on the properties of the galaxies in
the ENACS clusters: in particular their morphological types (inferred
from the spectrum) and their projected distribution. Combining the
morphological data with CCD photometry and internal velocity
dispersions, we will study the Fundamental Plane of early-type
galaxies. We discuss some preliminary results about the universality
of the Fundamental Planes in a few ENACS clusters. For a summary of
the results based on the velocity information obtained within ENACS,
we refer the reader to the companion review by Biviano et al. in this
volume.

\end{abstract}




\section{Introduction}

In the late 1980's a large amount of observational information had
been obtained about rich clusters of galaxies. However, even though
the quantity of the available data was quite impressive, the selection
of the clusters and the quality of the data in general was not
sufficiently homogeneous for statistical studies of the properties of
rich clusters {\em as a class}. For that reason an ESO Key-programme
was carried out in the late 1980's and the early 1990's, with the aim
to provide homogeneous velocity and magnitude data for a well-defined
sample of rich, nearby clusters. The cluster sample was chosen such
that, in combination with data in the literature, information would
become available for a complete, volume-limited sample of rich
clusters.

The observations of this survey, which has become known as the ENACS
(the ESO Nearby Abell Cluster Survey) have produced a catalogue of
redshifts and magnitudes for 5634 galaxies in the directions to 107
ACO clusters (and cluster candidates). In this review we recapitulate
briefly the properties of the ENACS and we summarize the results of
the analyses to date, as far as these do not involve the velocities of
the individual galaxies. I.e., we describe briefly the observational
set-up, the properties of the cluster sample, and some of the
properties of the cluster galaxies.

In order to allow a study of the distribution and kinematics of the
various types of galaxies, and possible differences between them, we
have used the ENACS spectra to derive a classification in terms of
early- and late-type galaxies, and we discuss some results of that
classification. As we find significant differences between the
projected distributions of the various types of galaxy in the ENACS,
we also analyze the projected distributions of the total galaxy
samples using the COSMOS catalogue. 

Finally, we discuss an ongoing project in which we use the ENACS
spectra to derive internal velocity dispersions of individual
galaxies. Using also photometric data from CCD imaging, we hope to
study the Fundamental Planes of the early-type galaxies in 20-25 ENACS
clusters. Here, we present some preliminary results from that project.

\section{The Survey}

The sample of target clusters constituting the ENACS was defined to
contain all R$_{\rm ACO} \ge$1 clusters in a 'cone' of 2.55 sr
centered more or less on the Southern Galactic Pole, which had a
spectroscopic redshift less than 0.1 or, failing that, a value of
m$_{10}$ in the ACO catalog (Abell et al. 1989) less than 17.0. The
latter condition implies a high probability for the redshift to be
less than 0.1. In the ENACS we obtained multi-object spectroscopy for
those clusters in this volume (the `cone') for which no extensive
redshift data were available in the literature.

\begin{figure}
\plotfiddle{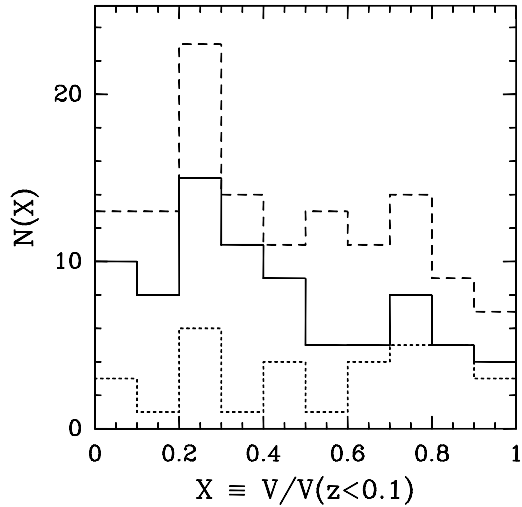}{6.6cm}{0}{50}{50}{-200}{-210}
\caption{The number of clusters in 10 equal-volume shells within a
redshift of 0.1, for all 128 clusters in the combined ENACS/literature
sample (dashed line), for the 80 clusters with at least 10 ENACS
redshifts (solid line), and for the richest 33 clusters (dotted
line).}
\label{f-1}
\end{figure}

In Fig.\ref{f-1} we show the redshift distribution of the combination
of ENACS and literature clusters in the `cone'. This figure gives the
number of clusters in 10 equal-volume shells between redshifts 0.0 and
0.1. So, for constant space density these numbers should be equal to
within the statistical noise. From this figure we conclude that the
ENACS plus literature clusters constitute a truely volume-limited
sample of R$_{\rm ACO} \ge$1 clusters. We refer to Mazure et al (1996)
for a more complete discussion of the volume-completeness of the
sample, and of the completeness with respect to richness.

The samples of galaxies that were observed spectroscopically were
defined from scanning with the Leiden Observatory plate-measuring
machine of, mostly, film copies of the SERC IIIa-J plates or glass
copies of the red PSS-I survey plates. A comparison with the COSMOS
catalogue shows the ENACS galaxy samples to be magnitude-limited, but
positionally unbiased subsets of the COSMOS catalogue. The effective
completeness limits of the samples of galaxies with redshifts range
from 16.5 to 17.5 in R$_{\rm 25}$, but redshifts were obtained
for a substantial fraction of the galaxies beyond these limits (see
Katgert et al. 1998).

As the name of the survey implies, the observations for the ENACS were
done with ESO telescopes at La Silla, viz. with the 3.6-m telescope
equipped with the OPTOPUS multi-object fibre-fed spectrograph for the
spectroscopy, and with the Danish 1.54-m and 0.92-m Dutch telescopes
for the calibration of the photometry through CCD-imaging. The
resolution of the multi-object spectroscopy was 130\AA/mm, and with
the 2.3 arcsec diameter fibres projecting onto $\approx$ 50$\mu$m on
the detector, the effective resolution was about 6\AA.

The spectroscopic observations consist of 170 CCD exposures each
containing about 50 spectra (see Katgert et al. 1996, for an example).
Effective integration times ranged from 60-100 minutes, often divided
up into two equal-duration exposures. Redshifts were obtained from
cross-correlation with template spectra, and much care was taken to
establish the linearity of the redshift scale and the correctness of
its zero-point. As a result, the advertized accuracies of velocities
of between 40 and 120 km/s are thought be realistic total (i.e. not
just internal) errors (see Katgert et al. 1996, for a discussion of
the reliability and robustness of velocity estimates in the ENACS).
Note that within the ENACS we found no evidence for the systematic
difference between redshifts from absorption and emission lines, as
reported recently by e.g. Cappi et al. (1998), who also used the
OPTOPUS multi-object system.

\begin{figure}
\plotfiddle{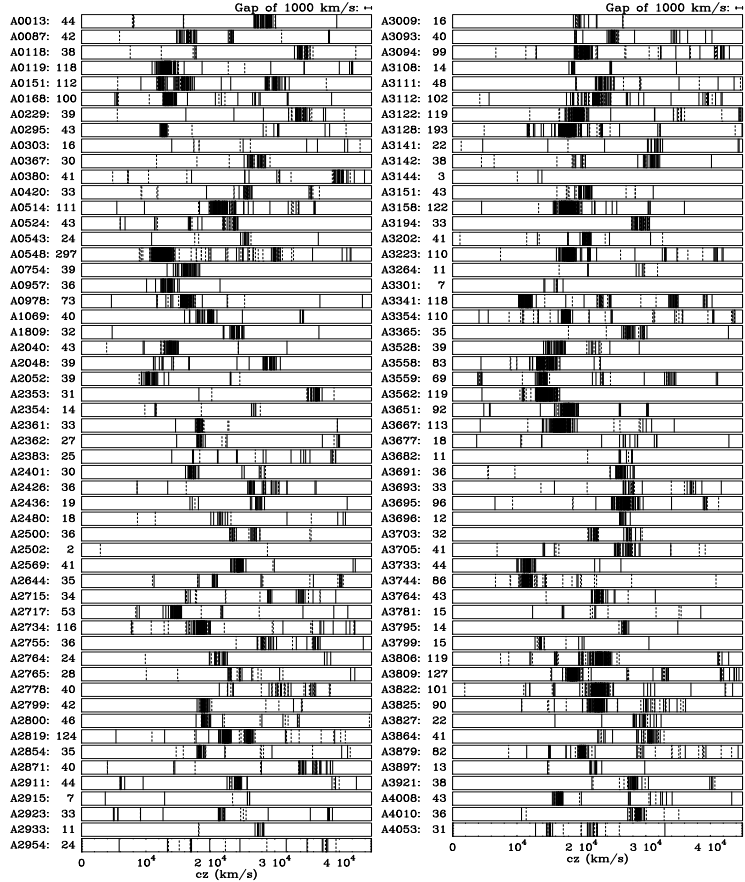}{12.5cm}{0}{50}{45}{-150}{-22}
\caption{Overview of the ENACS spectroscopic data, in the form of 107
`velocity bars', where each galaxy is indicated by a vertical line.}
\label{f-2}
\end{figure}

The spectral range of the observations was from $\approx$3900 to
$\approx$5600\AA\ which, for redshifts between 0.05 and 0.1 allows
detection of the emission lines OII 3727\AA, H$\beta$ 4860\AA\ and
OIII 4959/5007\AA. About 1 in 5 of the galaxies in the ENACS shows at
least one of these emission lines in its spectrum (see Biviano et
al. 1997). The fraction of galaxies with at least one of these
emission lines (ELG) decreases to about 1 in 8 if one considers only
the galaxies in compact velocity systems (i.e. clusters). About 75\%
of the total of 5634 galaxies in the ENACS is in such a system with at
least 4 members. Using morphological information from Dressler,
Biviano et al. (1997) concluded that 6 out of 7 of the ELG are
spirals, while the ELG represent about 1/3 of the total spiral
population.

In Fig.\ref{f-2} we show a summary of the redshift information
provided by the ENACS. This figure illustrates that the large majority
of the ACO clusters with z $\la$ 0.1 and R$_{\rm ACO} \ge$ 1 are real.
From a detailed analysis, Katgert et al. (1996) concluded that about
90\% of these peaks in the projected galaxy distribution correspond to
compact structures in redshift space.

The ENACS redshift catalogue has been made public and can be accessed
via http://cdsweb.u-strabg.fr/abstract.html; it is described in
Katgert et al. 1998 and Katgert et al. 1996.

\section{Galaxy populations in the ENACS clusters}
\subsection{Classification on the basis of the ENACS spectrum}

In addition to using the ENACS spectra for deriving redshifts by
cross-correlating with template spectra, we have recently used the
spectra also to estimate the type of a large fraction of the galaxies
in the ENACS catalogue (de Theije and Katgert, 1998). Note that the
selection of the galaxies for the spectroscopic observations was done
on survey plates with a scale that did not allow the determination of
the morphological type of the galaxy for all except the most extended
galaxies. Actually, we did not classify the galaxies at all in the
selection process, but fortunately morphological classification by
Dressler is available for a few hundred galaxies in the ENACS
(Dressler 1980b), and this information was used to calibrate our
spectral classification procedure.

\begin{figure}
\plottwo{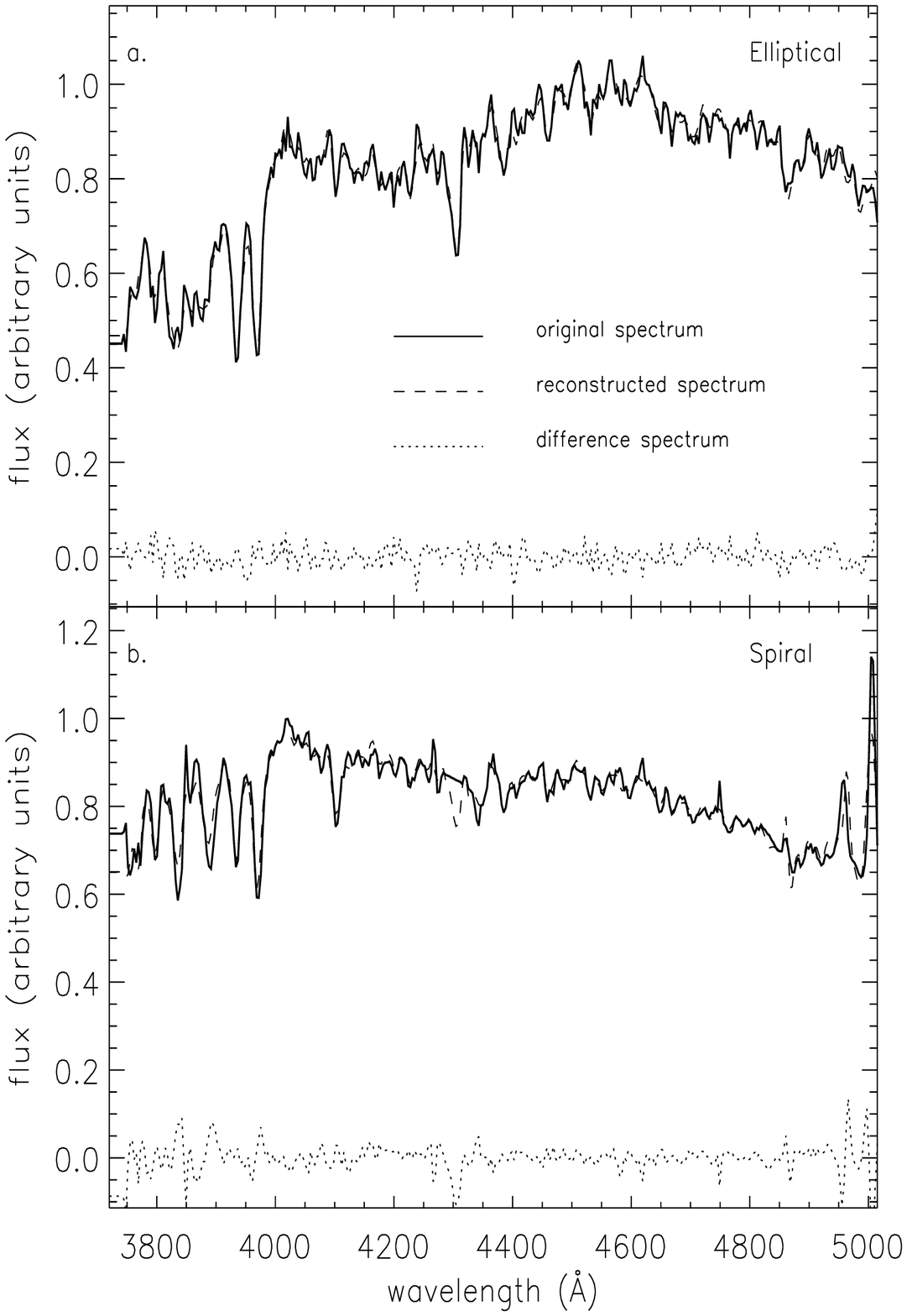}{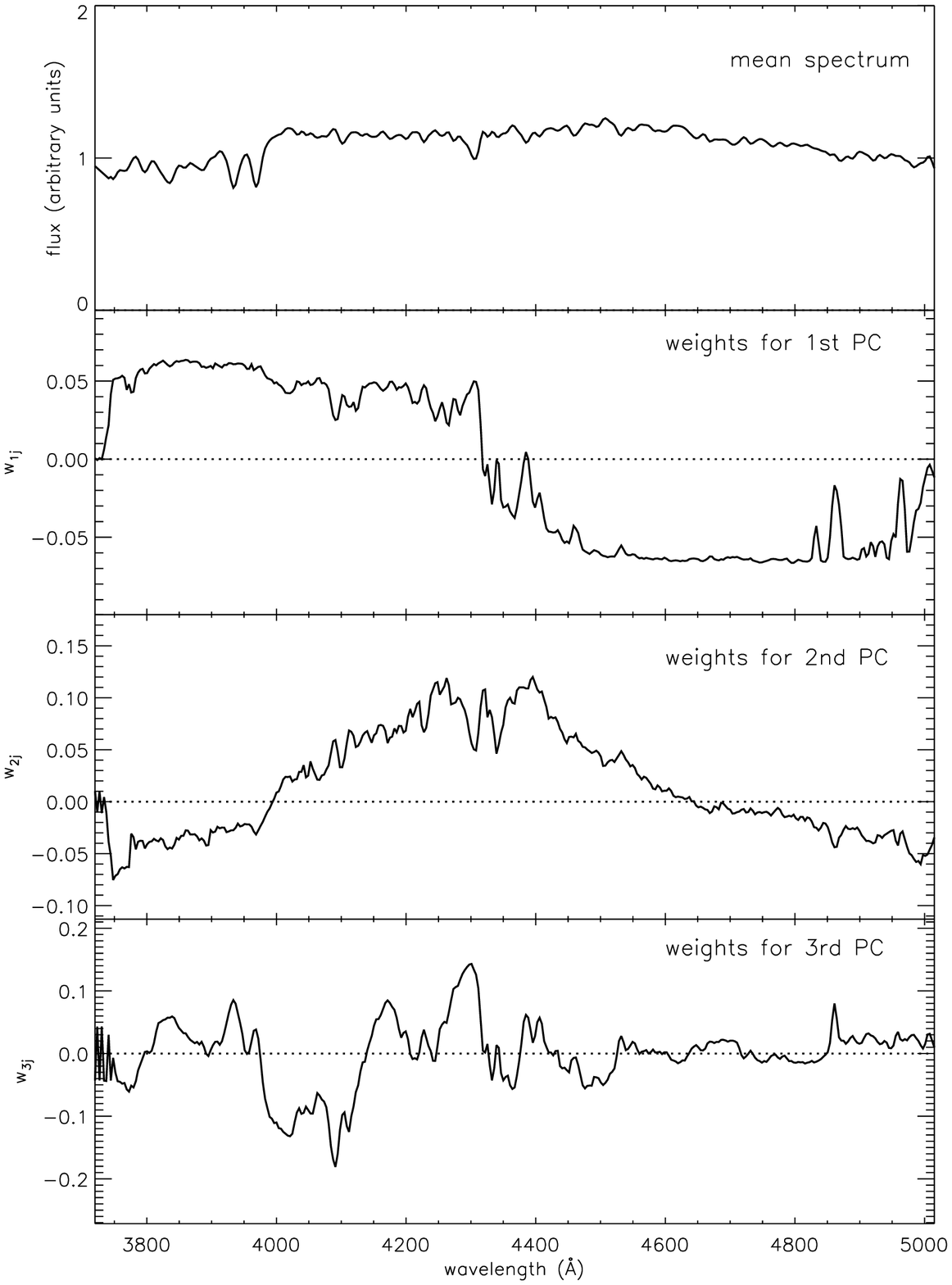}
\caption{Example of representation of 2 galaxy spectra by their first
15 Principal Components (left), and wavelength dependence of the
weights for the first three Principal Components (right).}
\label{f-3}
\end{figure}

In order to estimate the galaxy type from the spectrum we first
derived for each spectrum the 15 most significant Principal Components
(PCs), $e_1$ to $e_{15}$. In other words: the information in each
spectrum, which we first sampled with 371 spectral fluxes, was
condensed into 15 numbers. The latter cannot give a full description
of the spectrum, but they do very nearly so, as can be seen from the
left-hand panel in Fig.\ref{f-3}.

The physical interpretation of the weights $w_{ij}$, which define the
PCs $e_i$ as follows
\begin{displaymath}
\ e_i = \sum_j w_{ij} (f_j - <f_j>) /\sigma_j 
\end{displaymath} 
in which f$_j$ is the j-th flux value in the normalized spectrum and
$\sigma_j$ is the dispersion of that normalized flux value among all
spectra, can be gauged from the right-hand panel in Fig.\ref{f-3} for
$e_1$ to $e_3$. Clearly, $e_1$ is very much like a colour, as it
measures the slope of the continuum and $e_2$ measures its curvature.
The meaning of $e_3$ is somewhat less obvious although the 4000\AA\
break and the G-band seem to be involved. The meaning of the higher
PCs is even less obvious.

By themselves, the PC's do not provide a very good discrimination
between the spectra of galaxies of different morphological types,
although the value of $e_1$ increases systematically from early- to
late-type galaxies. However, the distributions of $e_1$ for the
galaxies with emission lines (which are mostly spirals, and which
constitute about one-third of all spirals) and those without emission
lines (early-type galaxies and the other two-thirds of the spirals)
suggests that a better separation may be possible if one uses the most
significant PCs together.

Therefore, the 15 PCs $e_1$ to $e_{15}$ were used as input for an
Artificial Neural Network (ANN), with 15 input values, one hidden
layer with 5 nodes, and a single continuous output node. The ANN was
trained with morphological types from Dressler (1980b). As the spectra
of ellipticals and S0's are quite similar, the separation between
these classes on the basis of the spectrum is not very good. For that
reason, a two-class system was used with an early-type (E+S0) class
and a late-type (Sp+Irr) class. The fact that most of the ELG are
known to be spirals was used to fine-tune the separation between
early- and late-type galaxies in the value of the output-node.

Using an independent set of galaxies from Dressler (1980b) to test the
performance of the combined PCA/ANN classification, we find that the
success rate is about 75\%, i.e. on average 1 out of 4 of the
announced 'spectroscopic' types does not agree with the morphological
type. A large part of these 'errors' must be due to an intrinsically
non-perfect spectral separation of early- and late-type galaxies, and
these are unavoidable and probably more or less symmetric between the
two classes. However, part of the 'errors' may be due to a real
inconsistency between spectrum and morphology, especially for spirals
which may have an early-type (bulge-) spectrum when observed in the
central 2--3 kpc region (corresponding to the diameter of the fibres
used for the ENACS spectroscopy). In this respect it is noteworthy
that for the separation between early- and late-type galaxies that we
adopted, 4 out of 5 of the early-type galaxies are classified
correctly from the spectrum, but only 2 out of 3 of the late-type
galaxies. This is the asymmetry one would expect for the type of real
inconsistency between morphological and spectral classification
mentioned above.

\subsection{Projected distributions of various types of galaxies}

\begin{figure}
\plotone{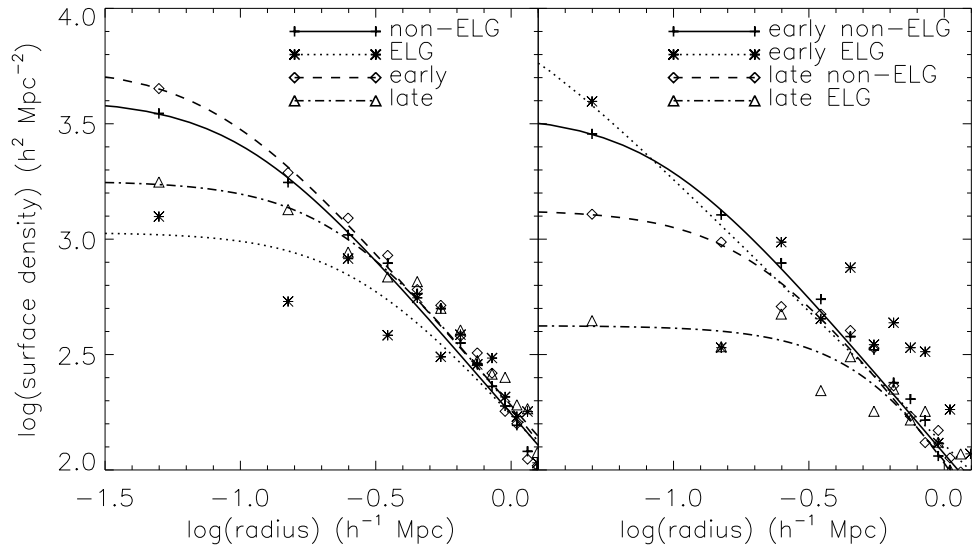}
\caption{Average projected number densities of the various types of
galaxies in the ENACS clusters (early/late and ELG/non-ELG).}
\label{f-4}
\end{figure}

The PCA/ANN analysis yielded `spectral types' for 3798 of the 5634
ENACS galaxies (as the PCA was done on a fixed zero-redshift interval,
spectra of galaxies with too high or low redshifts could not be
included, and for one of the observing periods, the calibration was
not sufficiently good). For the 3798 galaxies with spectral types we
also have information on the presense or absence of emission lines.
Using that information, we have derived the projected density profiles
for the various galaxy types, for a composite clusters of 66 ENACS
clusters containing 2594 galaxies, and those profiles are shown in
Fig.\ref{f-4}.

This figure confirms the morphology-density relation first quantified
by Dressler (1980a). It also confirms the differences between the
projected distributions of galaxies with and without detectable
emission lines, ELG and non-ELG respectively (see e.g. Biviano et
al. 1997). We now find that within the late-type class the galaxies
{\em without} emission lines are more centrally concentrated than
those {\em with} emission lines. Therefore, the late-type ELG avoid
the central regions of their cluster more strongly than any other type
of galaxy. For an interpretation of this result, kinematical data are
also required, for which we refer to Biviano et al. (1997, and this
volume) and de Theije and Katgert (1998).

\section{The overall projected galaxy density in the ENACS clusters}
\subsection{Cores or central cusps?}

In recent years, the mass profile of clusters has received renewed
attention, largely as a result of the work by Navarro, Frenk and White
(hereafter also NFW, e.g. 1997) who found that the density
distributions of dark matter halos in their simulations all follow a
universal profile, when the differences in mass are properly taken
into account. This universal profile has a clear cusp for which
observational confirmation has been sought in cluster mass profiles
based on observations. The experimental determination of mass profiles
requires application of the Jeans equation for the solution of which
both the number density profile of tracers of the potential (galaxies)
and their kinematics must be known.

However, barring a complete solution of the Jeans equation one can
investigate the presence or absence of a cusp in the projected number
distribution of galaxies. This type of analysis has a long history,
and until recently common wisdom held that galaxies (and probably also
total mass) in clusters follow distributions that have a core, like
the King and Hubble profiles. This view has been challenged, in
particular by the modelling of the mass distributions in clusters
which act as a gravitational lens (e.g. Kneib et al, 1993) and by the
result of NFW.

As the ENACS cluster sample is essentially volume-limited, it presents
a good starting point for an analysis of the galaxy distributions in
(rich) clusters. However, as the ENACS galaxy samples are subsets of
the total galaxy population in these clusters, we decided to use
galaxies from the COSMOS catalogue in apertures centered on those
ENACS clusters, which from the redshift data are known to be real. The
disadvantage of the COSMOS catalogue is that even towards the rich
ENACS clusters it contains a substantial contribution of unrelated
fore- and background galaxies, but the advantage is that completeness
to faint magnitude limits is ensured, so that many more galaxies are
included than in the ENACS samples.

We have analyzed the projected galaxy distributions in about 60
individual clusters, and in a composite cluster of 29 ENACS clusters
which from the COSMOS galaxy density contours are known to contain
little, if any, substructure. For the individual clusters, there is a
general tendency for the galaxy distributions to favour profiles with
a core rather than a cusp, but the preference is seldom very
significant. The evidence from the composite cluster, which contains
close to 5000 COSMOS galaxies, is more robust. Note that in the
construction of the composite cluster we took every care to avoid
destruction of a possible cusp (due to bad centering), as well as
creation an artificial cusp due to neglecting the ellipticities of the
individual galaxy distributions (see Adami et al. 1998).

\begin{figure}
\plotfiddle{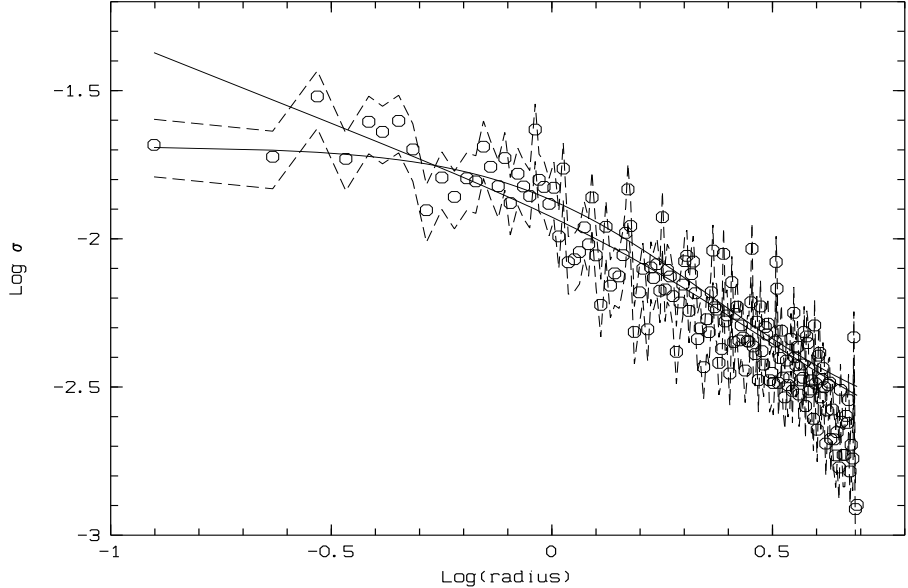}{7.8cm}{-90}{50}{45}{-200}{255}
\caption{The projected galaxy number density profile in a composite
cluster of 29 regular ENACS clusters (based on the COSMOS catalogue).
The open circles indicate the observed values, and the dashed lines the
1-$\sigma$ uncertainty range.}
\label{f-5}
\end{figure}

From Maximum-Likelihood fits to the galaxy distribution in the
composite cluster we obtained likelihood ratios which indicate a clear
and significant preference for profiles with a core rather than a cusp
(at significance levels of between 95 and 99 percent, depending on the
way the projected distances are scaled in the determination of the
composite cluster). This result is supported by a visual comparison of
the observed projected density distribution in the composite cluster
with a King- and a 2-D analogue of the NFW profile, as can be seen in
Fig.~\ref{f-5}.

From the likelihood ratios derived for subsets of the galaxies
selected on the basis of absolute magnitude, we conclude that the
preference for profiles with cores is strongest for the fainter
galaxies. The preference for profiles with cores is not observed for
the brighter one-quarter of the galaxies, but the latter do not show a
strong preference for profiles with cusps either.

\subsection{The outer slope of the density profile as a 
            cosmological probe}

In recent years, several authors have discussed the relation between
the outer slope of the density profiles of clusters on the one hand,
and the details of the formation process of large-scale structure in
the Universe on the other hand (e.g. Crone et al., 1994, and
references therein, Jing et al. 1995, to name but a few). Although a
description of the formation process includes the shape and the
amplitude of the initial fluctuation spectrum, it appears that the
value of $\Omega_0$ has a very strong influence on the average value
of the outer slope of the cluster density profile. Generally speaking,
the lower $\Omega_0$ is the steeper is the density profile, and this
global trend can be understood in terms of the dependence on
$\Omega_0$ of the amount of material `raining' in on the clusters at
the present time.

\begin{figure}
\plotfiddle{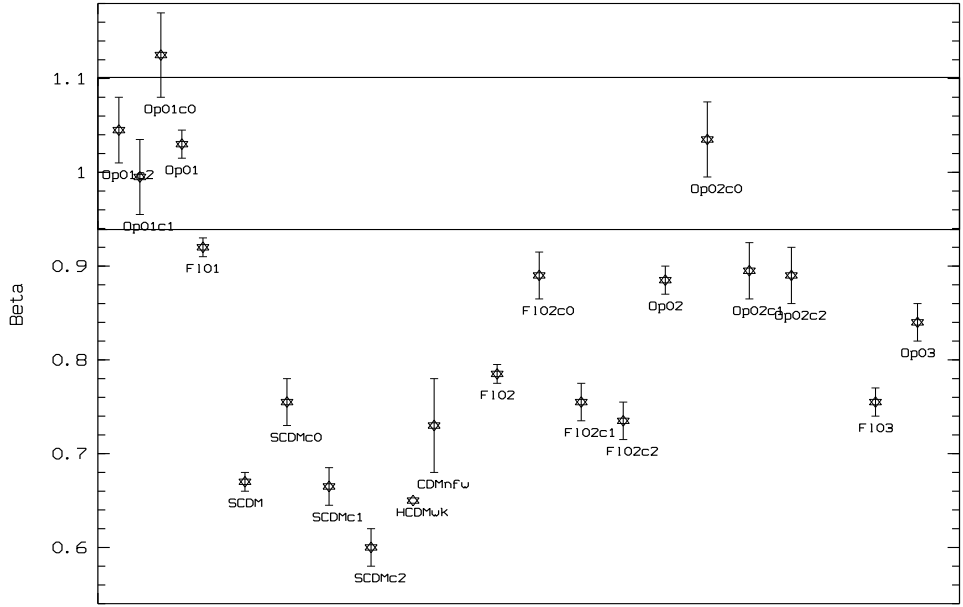}{8.0cm}{-90}{50}{45}{-220}{258}
\caption{The range of outer slope of the galaxy distribution
($\beta_{\rm 2D}$) allowed by our observations (horizontal lines at
0.94 and 1.10), compared with predictions from numerical simulations
taken from the literature.}
\label{f-6}
\end{figure}

For the individual clusters as well as for the composite cluster we
have made Maximum Likelihood fits to the galaxy distribution in which
we solved for position, elongation, characteristic radius, outer slope
and background (see Adami et al. 1998). In view of the preference for
profiles with cores we have fitted King profiles with a generalized
2-D outer slope, $\beta_{\rm 2D}$ as follows:
\begin{displaymath}
\Sigma(r) = \Sigma_0 \{ 1 + {(r/r_c)}^2 \} ^{- \beta_{\rm 2D}}
\end{displaymath} 
From these fits to the data we find an average value $\beta_{\rm 2D} =
1.02 \pm 0.08$. Using the relation $\beta_{\rm 2D} = \beta_{\rm 3D} -
0.5$ we have been able to compare our observations with average values
of $\beta_{\rm 3D}$ from numerical models for various cosmological
scenarios. The result is shown in Fig.~\ref{f-6}. For a detailed
description of the codes used to identify the scenarios, we refer to
Adami et al. (1998). Here we only summarize the main conclusion of
this comparison, which is that only the models with $\Omega_0$ of at
most a few tenths produce values of $\beta_{\rm 2D}$ that are within
the range allowed by the observations (note that direct estimates of
$\Omega_0$ through $M/L$-ratios of clusters give similar values).
Although not all possible scenarios have been included in the
comparison, this conclusion seems to be rather robust; there even
seems to be an indication that in general models with low $\Omega_0$,
but with a flat geometry (i.e. with $\Omega_{\Lambda} \neq 0$) do not
do a good job.

\section{The Fundamental Plane of Early-type Galaxies in ENACS 
         clusters}

As is well-known, the Fundamental Plane (hereafter FP) of early-type
galaxies is the relation between the two photometric parameters, $r_e$
and $\mu_e$, which describe the structure and luminosity of a galaxy,
and the central velocity dispersion due to the motions of the stars in
the galaxy. It was first studied by Dressler et al. (1987) and
Djorgovsky and Davis (1987). If light traced mass, and if all
early-type galaxies had the same (phase-space) structure, this
relation should simply reflect the virial equilibrium of the systems.
However, there are indications that the $M/L$-ratio of galaxies
increases somewhat with increasing luminosity, while there may be
differences in phase-space structure between different galaxies.

\begin{figure}
\plotfiddle{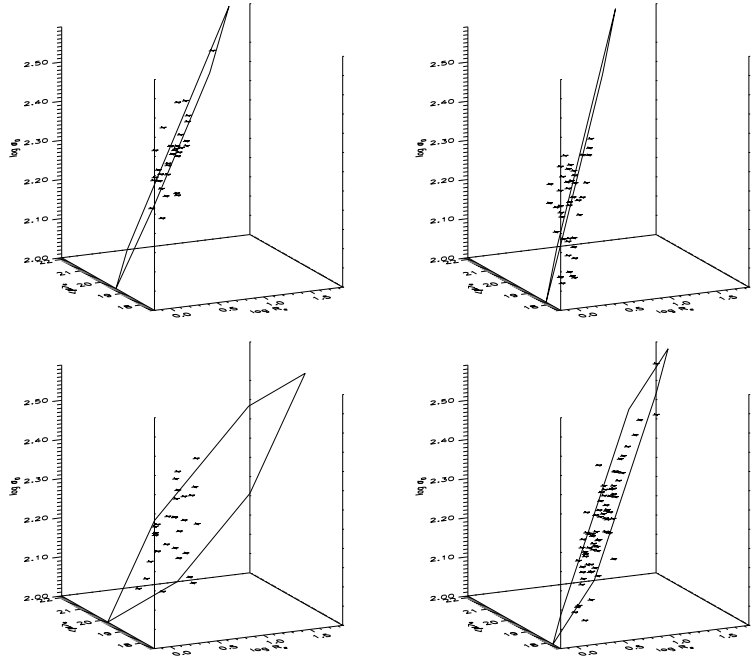}{8.2cm}{0}{60}{60}{-180}{-94}
\caption{Three-dimensional representations in log$\sigma$ (vertical)
vs. log r$_e$ and log $\mu_e$ space of the Fundamental Planes of
early-type galaxies in three ENACS clusters and in Coma (lower
right).}
\label{f-7}
\end{figure}

Nevertheless, J\"orgensen et al. (1996) find that the early-type
galaxies in the Coma cluster define a very narrow FP, and that the
FP's of various local clusters are consistent with the assumption of a
universal FP, although this conclusion is not very strong in view of
the sometimes rather limited statistics. Recently, the FP's in 4
clusters at different redshifts out to z = 0.58 were shown to be
consistent with a universal FP, with a redshift dependence that is
consistent with passive luminosity evolution of the early-type
galaxies (Kelson et al, 1997). Although all evidence thus seems
consistent with the assumption of a universal FP, we have embarked on
a study of the FP's in about 20 rich, nearby ENACS clusters.

The photometric parameters that are required for the determination of
the FP are obtained through CCD-imaging with the Dutch 0.92-cm
telescope at La Silla. At present, close to 1500 images have been
obtained, about 850 of which are of early-type galaxies. The velocity
dispersions are derived from the ENACS spectra, calibrated with
long-slit observations. Preliminary results appear to indicate that,
when properly analyzed, the ENACS spectra can yield reliable estimates
of the internal velocity dispersions for about half of the galaxies in
the ENACS catalogue. In Fig.~\ref{f-7} we show the provisional results
for three ENACS clusters and the Coma cluster (lower right). It is
clearly too early to reach firm conclusions, but there may be an
indication that there are differences between the various FP's, either
in orientation of the FP, its flatness or the dispersion around it.

\section{Conclusions}

We have summarized the main characteristics of the ESO Nearby Abell
Cluster Survey (the ENACS), and described some of the recent results
obtained from it. The ENACS defines, in combination with literature
data, a volume-limited sample of R$_{\rm ACO} \ge$1 clusters, with
good redshift and magnitude data. We have used the ENACS spectra to
estimate the morphological type of the galaxies. Using these spectral
'morphological' types, we studied the distribution of early- and
late-type galaxies in clusters. The late-type galaxies appear to be
less centrally concentrated than the early-type galaxies, and in
particular the late-type galaxies with emission lines in their spectra
appear to avoid the central regions. Using COSMOS data for a subset of
regular ENACS clusters we have studied the projected galaxy
distribution, and we find no evidence for the cusp that is expected
from the Navarro, Frenk and White density profile. The outer slope of
the projected galaxy distribution seems to indicate a fairly low value
of $\Omega_0$.  Finally, we show some preliminary results on the
universality of the Fundamental Plane of early-type galaxies in
clusters.

\acknowledgments

We thank ESO for the allocation of the observing time. We acknowledge
contributions of other members of the ESO Cluster Key Programme, in
particular those of Roland den Hartog and Jaime Perea.


%

\end{document}